\documentclass[11pt]{article}
\usepackage{moriond03}

\def\Journal#1#2#3#4{{#1} {\bf #2}, #3 (#4)}

\def\NPB{{\em Nucl. Phys.} B}
\def\NPA{{\em Nucl. Phys.} A}
\def\PLB{{\em Phys. Lett.}  B}
\def\PRL{\em Phys. Rev. Lett.}
\def\PRD{{\em Phys. Rev.} D}

\def\be{\begin{equation}}
\def\ee{\end{equation}}
\def\bea{\begin{eqnarray}}
\def\eea{\end{eqnarray}}

\newcommand{\xbj}{x_{\scriptscriptstyle B}}
\newcommand{\text}{\hbox}
\newcommand{\amp}[1]{\langle #1 \rangle}
\newcommand{\bm}[1]{\mbox{\boldmath $#1$}}
\begin{document}
\vspace*{4cm}
\title{REVIEW OF QCD SPIN PHYSICS}

\author{  DANI\"{E}L BOER           }

\address{Department of Physics and Astronomy, Vrije Universiteit Amsterdam\\
De Boelelaan 1081, NL-1081 HV  Amsterdam, The Netherlands}

\maketitle\abstracts{
A short review is given of QCD spin physics and its major aims: obtaining the 
polarized gluon density, the transversity 
distribution and understanding single spin asymmetries. 
The importance of the Drell-Yan process, the role of electron-positron 
colliders and the use of 
polarization to probe other, not spin specific physics are emphasized. 
}

\section{Polarized structure functions and parton densities}

The polarized structure functions $g_1$ and $g_2$ of Deep Inelastic Scattering
(DIS) of polarized electrons off polarized protons (or other spin-$1/2$
hadrons), i.e.\ $\vec{e} \, \vec{p} \to e' \, X$, appear in
the parametrization of the hadronic part of the cross section, given by the 
hadron tensor
\be
{W_A^{\mu\nu}}
={\frac{i\epsilon^{\mu\nu\rho\sigma}q_{\rho}}{P\cdot q} }\, 
\left[
{S_\sigma} {g_1(\xbj,Q^2)} + {
\left( S_\sigma - \frac{S\cdot  q}{P\cdot  q}
P_\sigma \right)} {g_2(\xbj,Q^2)} \right],
\ee
with hadron momentum $P$ and spin vector $S$, photon momentum $q$,  
$\xbj = Q^2/2P\cdot q$ and $Q^2 = -q^2$. 
The definition of {structure functions} is independent of 
the constituents of the hadron. The {pQCD improved parton model} allows one 
to go to the quark-gluon level, such that 
the polarized structure functions are 
expressed in terms of parton distribution functions. This exemplifies the 
goal of QCD spin physics, namely to understand the spin structure of 
hadrons in terms of quarks and gluons. For the longitudinal spin or {helicity}
the (leading twist) parton distributions are $\Delta q, \Delta \bar q, 
\Delta g$ and for {transverse spin} $\delta q, \delta \bar q$ 
($\delta g = 0$ due to helicity conservation).

One sub-goal is to complete the {spin sum rule}. The sum of the contributions 
to the proton spin have to add up to 1/2:
\be
\frac{1}{2} = \frac{1}{2} \Delta \Sigma + \Delta g + L_z \qquad 
\left( = \frac{1}{2} \Delta \Sigma + {L_q + J_g} \right),
\ee
where $\Delta \Sigma = \Delta u + \Delta d + \Delta s$ is the total
contribution of
the spin of the quarks, $\Delta g$ that of the gluons and $L_z$ is the orbital
angular momentum of the quarks and the gluons together ($L_q$ stands for the
orbital
angular momentum of the quarks alone and $J_g$ for the total angular momentum
of the gluons). A still open problem is whether {$L_z =L_q + L_g$}?  
(or equivalently, {$J_g = \Delta g + L_g$}?) 
Here one requires that the individual quantities 
should be separately measurable and defined in a gauge invariant, process 
independent way, cf.\ Ref.\ \cite{Jaffe} and references therein.

Only lightcone momentum fraction ($x$) integrated information enters in the 
sum rule: $\Delta q = \int_0^1 dx \left[ (q_+ - q_-) + (\bar q_+ - \bar q_-)
\right]$ and $\Delta g = \int_0^1 dx \left[ g_+ - g_- \right] $ (where $\pm$
stands for the helicity of the proton).  
Experiments find that $\Delta \Sigma \sim 0.3$ and $\Delta s \sim
{-} 0.1$. Such statements ought to be accompanied by the renormalization
scheme and scale at which these numbers hold, but here we only want to mention
that such small numbers for $\Delta \Sigma$ and such relatively 
large values 
of $\Delta s$ were completely unexpected and viewed as a `spin crisis' or 
`spin puzzle'. In the
near future the $\Delta g$ piece of the puzzle will be determined 
experimentally and then the relative 
importance of the orbital angular momentum is determined implicitly.  

But of course {one is not only interested in the decomposition of $1/2$.
One also wants an accurate description of $\Delta q(x), \Delta g(x), \delta
q(x)$ as they appear in processes like DIS, Drell-Yan, etc, namely as function
of $x$. In other words, one wants to obtain a {complete map of the 
spin structure of the proton} (as function of $x$ and $Q^2$). 
At {leading twist} this entails: knowing $\Delta q(x)$ and $\delta q(x)$ 
for all quark flavors and knowing $\Delta g(x)$}.

From inclusive DIS, to be specific, from the structure function $g_1(x)$, 
one can only get $\Delta q(x) + \Delta \bar q(x)$ information 
and  $\Delta g(x)$ only implicitly via evolution, 
using
\be
g_1^{p/n} =  
\left(\frac{1}{9} \Delta \Sigma \pm \frac{1}{12} \Delta q_3^{{\rm
NS}} + \frac{1}{36} \Delta q_8^{{\rm NS}} \right) \otimes \left(1 +
\frac{\alpha_s}{2 \pi} \Delta C_q \right)
+ \sum_q e_q^2 \frac{\alpha_s}{2\pi}
\Delta g \otimes \Delta C_g, 
\ee
hence other processes are needed.
At {\footnotesize RHIC (BNL)} 
polarized $p \, p$ collisions will be performed, in order to 
measure $\Delta g(x)$ 
and $\Delta \bar q(x)$ in a variety of ways (for $\delta q(x)$ see below). 
For $\Delta g(x)$ one
can study $\vec{p} \, \vec{p} \to \gamma \, X; \, \, {\rm jet} \, X; \, \, 
\gamma \, {\rm jet}; \, \, \text{jet} \, \text{jet}; \, \, \pi^0 \, X$; 
$c\bar c\, X; \, \, b \bar b\, X; \ldots$. For $\Delta q(x), 
\Delta \bar q(x)$ one can study {$\vec{p} \, p \to W^\pm \,  X $}. 
In addition, there will be more (semi-)inclusive DIS data from 
{\footnotesize COMPASS (CERN), HERMES (DESY)} and {\footnotesize JLAB}.

The structure function $g_2$ has also been measured ({E155} 
Collaboration at
{\footnotesize SLAC} \cite{E155}), 
albeit with much less precision than $g_1$. The combination
$g_1 + g_2 = g_T$ contains information about quark-gluon correlations inside
the proton's transverse spin. This is a higher twist effect and $g_T$ is not 
related to the leading twist, transverse spin (i.e.\ helicity flip) parton 
density $\delta q$.

\section{Transversity}

Transversity ($\delta q $) is completely unknown (no data). It cannot be 
measured in inclusive DIS (heavily suppressed). 
The reason is that it must be 
probed together with another helicity flip. There are two main routes to
follow. The first is to use two transversely polarized hadrons, e.g.\
study $p^\uparrow \, p^\uparrow \rightarrow \ell \,
\bar\ell \, X$, $p^\uparrow \, p^\uparrow \rightarrow \, 
\text{jet} \, X$, $e \, p^\uparrow \to 
\Lambda^\uparrow \, X$ or $p \, p^\uparrow \to 
\Lambda^\uparrow \, X$. The second route is to use the distribution of final 
state hadrons, which may be correlated with the transverse spin direction. For
example, one can measure the transverse momentum of a final state hadron
compared to the jet direction. 
The so-called ``{Collins effect}'' \cite{Collins-93b} 
may correlate this transverse
momentum with the transverse spin and may produce single spin asymmetries in 
{$e \, p^\uparrow \to e' \pi \, X$} and 
{$p \, p^\uparrow \to \pi \, X$}.
Or one can measure the angular distribution of hadron pairs, where their
orientation may be correlated with the transverse spin, described by
the so-called
two-hadron {interference fragmentation functions}
\cite{Ji-94,ColHepLad,JJT,Bianconi} and leading to asymmetries in 
{$e\, p^\uparrow$ or $p\, p^\uparrow \to (\pi^+ \pi^-) \, X$}.

Several of these options contain {unknown fragmentation functions} that have
to be determined separately. For this purpose 
one can use {off-resonance data of 
$B$-factories} \cite{MGP}, such as {\footnotesize BELLE} or {\footnotesize 
BABAR}.
This would also be useful for the study of the {spin structure of 
hyperons}.

\section{Spin asymmetries in hadron and lepton pair production}

As said, the direction of produced hadrons may be correlated with the 
polarization of one or more particles in the collision. This is demonstrated 
by the large single spin asymmetries that have been observed in 
{$p\,  p^{\uparrow} \rightarrow \pi \, X$} \cite{Adams,AGS,Rakness}. 
It is a so-called left-right asymmetry, since the pions prefer to
go left or right of the plane spanned by the beam direction and the transverse
spin, depending on whether the transverse spin is up or down and depending on
the charge of  the pions.
Similar types of asymmetry have been observed in 
{$p \, p \rightarrow \Lambda^{\uparrow} \, X$} \cite{Bunce}
and {$\nu_\mu \, p \to \mu \, \Lambda^{\uparrow} \, X$} \cite{NOMAD}. 
It is expected that the
underlying mechanisms of these different asymmetries are related, but 
it is also fair to say that single transverse spin
asymmetries are {not really understood, i.e.\ 
it is not clear how to explain them on the quark-gluon level. 
The suggested mechanisms
can be roughly categorized as: semi-classical models; 
$\bm{k}_T$-dependent distributions; higher twist. 

One particularly informative observable is the single transverse spin 
asymmetry $A_N$ in Drell-Yan $p \, p^\uparrow \to \ell \, \bar \ell \, X$, 
since mechanisms that depend solely on 
fragmentation effects do not contribute. To indicate that an experiment with 
a (few) percent accuracy can be extremely useful to narrow down the possible
origins of single spin asymmetries, three
predictions are summarized, each based on a fit to the same 
E704 $\pi$ asymmetry data \cite{Adams} (no quantitative comparisons of the
predictions are possible however, due to the different kinematics chosen). 
\begin{itemize}
\item A {semi-classical model} calculation \cite{Boros} 
predicts for positive $x_F$ an asymmetry that starts out at +15\% at $x_F=0$ 
and grows quickly to +40\% for large $x_F$. This is for an invariant mass $Q$ 
of the lepton pair of 4 GeV and the asymmetry is slightly larger for $Q=9$ 
GeV (both at $\sqrt{s}=20$ GeV). The asymmetry is
still appreciable in size for small, negative $x_F$. 
The transverse momentum of
the lepton pair was partly integrated over. 
\item A recent
calculation \cite{Mauro} 
using the $\bm{k}_T$-dependent Sivers effect distribution function \cite{s90}
predicts (at $\sqrt{s}=200$ GeV) an
asymmetry that is negligible for $x_F<0.1$ and then grows in magnitude 
to become {\em minus\/} 
10-30\% for $6<Q<10$ GeV and $10<Q<20$ GeV, respectively, $|y|<2$, and 
at a particular fixed $q_T$ of the 
lepton pair that maximizes the asymmetry. The study nicely shows that
$\bm{k}_T$ dependence does not imply $1/Q$ suppression.
\item A higher twist prediction \cite{DB-Qiu} using the Qiu-Sterman
mechanism \cite{QS-91b} 
yields $|A_N^{DY}| \sim {70 \; {\rm MeV} /Q}$, e.g.\ 
${2\% \; \; \text{at}\;  \; Q=4 \; \text{GeV}}$ ($q_T$ integrated). 
The power law decrease with $Q$ is a distinctive feature 
of higher twist. The predicted asymmetry is approximately $x_F$ independent.
\end{itemize}

\section{Azimuthal spin asymmetries} 

Apart from the left-right asymmetries, {azimuthal} spin asymmetries 
have been observed. In semi-inclusive DIS (${e \, 
p \to e' \, \pi \, X}$)
the {\footnotesize HERMES} Collaboration \cite{HERMES} has measured a
nonzero {$\sin \phi$ asymmetry in $e \, \vec{p}$} scattering ($A_{UL}$). 
It is a 
2\% asymmetry for $\pi^+$. Soon there will also be data from {\footnotesize 
HERMES} on 
{$e \, {p}^\uparrow$} scattering ($A_{UT}$).   
The {\footnotesize CLAS} Collaboration (Jefferson Lab) has also observed \cite{CLAS} a  
nonzero {$\sin \phi$}, but in $\vec{e} \, {p}$ scattering ($A_{LU}$). 
These DIS data are at low $Q^2$ ($\amp{Q^2} \sim 1-3$
GeV$^2$), so the interpretation of the asymmetries is far from clear. But
they do demonstrate nontrivial spin effects, possibly related to
the asymmetries of the $p \, p$ experiments. 

\section{Spin as a tool}

An advantage of polarization is that Standard Model contributions 
(or at least QCD contributions) may be filtered out. One can for
instance study parity violation in polarized $p\,p$ scattering. Asymmetries in
the processes {$\vec{p} \, p \to \text{jet} \,  X $} 
(${A_L^{\, \text{jet}}}$) and {$\vec{p} \, \vec{p} \to \text{jet} \, X $} 
(${A_{LL}^{PV}, \, \bar A_{LL}^{PV}}$) defined as:
\be
{A_L^{\, \text{jet}}} = \frac{\sigma_{-} - \sigma_{+}}{\sigma_{-} + 
\sigma_{+}}, \qquad
{A_{LL}^{PV}} =\frac{\sigma_{--} - \sigma_{++}}{\sigma_{--} + \sigma_{++}}, 
\qquad {\bar A_{LL}^{PV}} = \frac{\sigma_{-+} - \sigma_{+-}}{\sigma_{-+} 
+ \sigma_{+-}},
\ee
are measures of parity violation. 
Similarly for {CP violation}: 
since the quark coupling to the $W$ has fixed helicity, certain 
transverse spin (helicity flip) asymmetries should be absent. For example, 
in {${p}^\uparrow \, p^\uparrow \to W \,  X$}, which may be
relevant for {\footnotesize RHIC} (upgrade) or a polarized {\footnotesize
LHC}. 

Another example where spin can be used as a tool to study other, not spin
specific physics is in the study of small $x$ effects. 
Polarization may offer new probes of gluon {saturation}. Asymmetries involving
polarization dependent (${\bm k}_T$-odd) functions can be sensitive to the
saturation scale $Q_s$. Recently, this was investigated theoretically for 
$p \, A \to \Lambda^\uparrow \, X$ \cite{DB-Dumitru}. 

\section{Conclusions}

Much experimental and theoretical work has been done on QCD spin physics 
over the last decades, but due to the complicated nature of QCD
the understanding of the spin of the proton is not yet complete
(let alone that of the neutron, $\Lambda$, $\rho$, etc). 
{Accurate determinations of polarized parton distribution and 
fragmentation functions} as function of $x$ and $Q^2$ are {still in 
progress}. Experimental efforts to measure new spin observables 
are  under way} at several laboratories ({\footnotesize BNL, CERN, DESY, JLAB,
SLAC, ...}).  

Striking single spin asymmetries have been observed (left-right
asymmetries and $\sin \phi$ azimuthal asymmetries), but are {still 
not understood} (pQCD and collinear factorization are insufficient).
Especially the single transverse spin asymmetry $A_N^{DY}$ in Drell-Yan
offers a good opportunity to learn about the underlying mechanism(s).

\section*{Acknowledgments}
I would like to thank the organizers of this stimulating meeting for their 
kind invitation. The research 
of D.B.~has been made possible by financial support from the Royal
Netherlands Academy of Arts and Sciences.

\end{document}